\documentclass{pasa}%

\usepackage[square,numbers]{natbib}
\usepackage{optprog}
\usepackage{graphicx}
\usepackage[normalem]{ulem}

\title[Almost Exact Portfolio Risk Budgeting with Return Forecasts for Portfolio Allocation]{Almost Exact Risk Budgeting with Return Forecasts for Portfolio Allocation}

\author[]{Avinash Bhardwaj$^{1,2}$, Manjesh K Hanawal$^1$, Purushottam Parthasarathy$^{1,*}$,   
\affil{$^1$IEOR, Indian Institute of Technology, Bombay }
\affil{$^2$Mechanical Engineering, Indian Institute of Technology, Bombay}
\affil{$^*$Corresponding author: 194192001@iitb.ac.in}

}%

\usepackage{aas_macros}
\usepackage{hyperref} 
\hypersetup{colorlinks,citecolor=blue,linkcolor=blue,urlcolor=blue}

\hypersetup{draft}
\newcommand{\R}{\mathbb{R}}

\begin{document}

\begin{frontmatter}
\maketitle

\begin{abstract}
In this paper, we revisit the portfolio allocation problem with designated risk-budget [Qian, 2005]. We generalize the problem of arbitrary risk budgets with unequal correlations to one that includes return forecasts and transaction costs while keeping the no-shorting (long-only positions) constraint. We offer a convex second order cone formulation that scales well with the number of assets and explore solutions to the problem in different settings. In particular, the problem is solved on a few practical cases - on equity and bond asset allocation problems as well as formulating index constituents for the NASDAQ100 index, illustrating the benefits of this approach.   
\end{abstract}

\begin{keywords}
risk-parity portfolio-optimization risk-budgeting mean-variance tactical-asset-allocation portfolio-allocation
\end{keywords}
\end{frontmatter}

\section{INTRODUCTION }
\label{sec:intro} In portfolio allocation problems, risk-budgeting and risk-parity are two important criteria that are closely related to each other. Although the investment management community is quite familiar with risk-parity as a concept, the term risk-budgeting has been less heard and talked about, both in academic and practitioner circles. In fact, it may come as a surprise that the term risk-budgeting, mentioned in \cite{arnott2002risk} actually predates the term risk-parity that was coined around the same time as \cite{qian2005financial}, where the authors provided a clear definition of risk contributions to a portfolio. Specifically, 
the problem of risk budgeting the portfolio was defined as the following - 
{\it given covariance information on a basket of assets, risk budgeting seeks to form a portfolio whose partial risks are weighted as per a pre-determined scheme.}

Consider $n$ assets indexed by set $[n]=\{1,2,\ldots,n\}$. Let $b_i \in [0,1]$ to be the fractional risk budget of asset $i \in [n]$ and
$C \in  \R^n \times \R^n$ their covariance matrix. Let $x \in\R^n$ representing the portfolio's fractional composition. The {\em partial risk} of asset $i$ under a variance risk measure can be expressed as $x_i (C x)_i$. Then, the {\em risk budgeting portfolio} for a specific risk measure, namely the variance, satisfies :
\begin{equation}\label{eq:par-def}
x_i(Cx)_i=b_i \,x^\top Cx,\quad i \in [n].
\end{equation}
Additionally, since the portfolio is fully invested and fractional risk budgets must add to 1 we have
\begin{equation}\label{eq:summation}
\sum_{i=1}^n x_i = 1 \mbox{ and } \sum_{i=1}^n b_i = 1 
\end{equation}
Summarizing the above observations, we note that the risk budget problem $(\mathbf{P})$ can be formulated as:
\begin{optprog*}\label{eq:rba-prog0}
minimize & \objective{ \sqrt{x^\top Cx}} \label{objective1}\\ 
$(\mathbf{P})$ \qquad subject to & x_i(Cx)_i & = & b_i\,x^\top Cx, & i \in [n] \label{RB} \\ 
& 1^\top  x & = & 1 \\
& x & \geq & 0
\end{optprog*}

It should be noted that the risk parity problem is a specific case of the risk budgeting problem where all fractional risk budgets are equal, i.e. $b_i = \frac{1}{n}$ $\forall\, i \in [n]$. In this work we re-formulate the generalized risk-budgeting problem with return forecasts and transaction costs as a min-risk budgeting problem (a second order cone program) and discuss the applications of this result to systematic asset allocation. 

\section{PRIOR WORK AND CONTRIBUTION}

\subsection{Prior Work}
As the concept of risk-parity took hold in the investment community through the 2008 banking crisis, \cite{qian2011risk} made a compelling case for risk-parity as an allocation strategy by focusing on the large risk allocation that popular 60-40 portfolios gave to equity markets. 
\cite{bruderroncalli2012} defined the risk budgeting problem as a general case of the risk parity problem and presented theoretical results on the variance of the resulting portfolio - that it is in between the minimum variance and the corresponding weight budgeting portfolio. The authors also analytically solved the problem for the two-asset case and presented existence and uniqueness results for the general case. The authors in \cite{roncalli2016risk} were able to extend the risk budgeting approach to risk factors - as an illustration, they showed that this approach can be used to allocate risk to the Fama-French factors in a systematic way.
In a work that seeks to understand the fundamental workings of risk-parity, \cite{asness2012leverage} proposed leverage aversion as a plausible reason as to why the average investor does not hold the risk parity 
portfolio. The authors also pointed out that not all investors have access to leverage, however, some do. These more sophisticated investors can
indeed benefit from the superior risk adjusted returns of the levered risk parity portfolio.  
Focusing more on work that makes computational advances towards calculating the weights in the risk-parity portfolio, \cite{MausserRomanko2014} reviewed existing formulations of the risk-parity portfolio (ERC - equal risk contribution portfolio), compared the empirical efficiency of solving this problem using a variety of techniques and proposed an alternate formulation that relied upon converting a hyperbolic constraint to a second order cone constraint. Consequently, they showed that the ERC portfolio with non-homogeneous correlations across assets can be solved as a second order cone program. 
\cite{cesarone2017equal} showed that the ERB (Equal Risk Bounding) is a superior technique than ERC for portfolio selection. In the case where short selling is allowed, the ERB portfolio was shown to be the same as the ERC portfolio. 
In \cite{mausser2018long}, the 2014 ERC portfolio SOCP formulation was extended to equal CVaR contributions.
\cite{gambeta2020risk} presented a formulation of the ERC portfolio that was relaxed to deviate from the ERC allocations to incorporate asset forecasts.
In more recent work, \cite{ANIS2022} solved the ERC portfolio with a cardinality constraint. The authors empirically demonstrate that these portfolios show good out of sample performance.

In this work we extend the results of \cite{MausserRomanko2014} to formulate an arbitrary risk budget portfolio with unequal correlations, return forecasts, transaction costs, and potentially also position constraints (the most generic case in portfolio optimization). We present a second order cone reformulation of the proposed problem and provide a computational analysis to demonstrate the efficacy and efficiency of the reformulation for examples with a large ($\approx$ 100) number of assets. 

\subsection{Problem Setup And Contribution}
We consider a set of $n$ assets. We further define by $C \in \R^{n \times n}$ and $r \in \R^n$, the 
positive-definite
covariance matrix and the vector of expected returns for these assets, respectively. We denote by $s$ a vector of investment sizes (denominated in \$) of a long-only portfolio and $x_k$ the corresponding fractional holdings in the $k^{th}$ asset:
\begin{equation} 
x_k = \frac{1} { \Sigma_i s_i}  s_k
\end{equation}


As shown by \cite{roncalli2013introduction} and \cite{spinu2013algorithm} in prior work, a 
solution to the risk budget problem  $(\mathbf{P})$ can be computed by solving an alternate problem $(\mathbf{P^*})$ given as follows:
\begin{optprog*}\label{eq:rba-progalt}
${\displaystyle \min_{x\in\R^n_+}}$ & \objective{ \frac{1}{2}x^\top Cx - \sum_{i=1}^n b_i \log x_i  } \label{objective2}
\end{optprog*}

As a brief explanation of why this works, observe that the first order optimality conditions for $(\mathbf{P^*})$ are
\begin{equation}
    (Cx)_i - \frac{b_i}{x_i}=0, \quad i\in[n],  \label{eq:KKT}  
\end{equation}
which are exactly the risk budgeting conditions in $(\mathbf{P})$ if we set the total variance of the portfolio in \eqref{eq:par-def}, without loss of generality, to 1. Note that the final portfolio satisfying the summation constraint \eqref{eq:summation} can be obtained by re-scaling the weights so they sum to 1. It is worth-while to note that this works as equation \eqref{eq:par-def} is scale-invariant - if a solution $x^*$ satisfies \eqref{eq:par-def}, so does $kx^*$. Problem $(\mathbf{P^*})$ is usually solved using some variant of Newton's method or block coordinate descent \cite{spinu2013algorithm}.


Further observe that any additions of more parameter(s) to the objective function in the original problem (adding asset return forecasts or accounting for transactions cost) or adding additional constraints will alter the first order optimality conditions \eqref{eq:KKT}. Consequently, target risk budgets may not be achieved. In other words, the convex formulation above only works in a very specific (almost impractical) case - with no return estimates, no  position constraints, or transaction costs.
It is noteworthy that \cite{MausserRomanko2014} use a different approach in formulating the ERC portfolio, a special case of the risk budgeting portfolio, as an SOCP program. Their approach could potentially handle the extra constraints that are proposed in this work, however the formulation is specifically for computing an ERC portfolio.

Our work extends the work of \cite{MausserRomanko2014} to show the generalized second order cone program formulation for an arbitrary risk budgeting portfolio allocation with return forecasts and transaction costs. We solve arbitrary risk budgeting exactly and argue its merits as a portfolio allocation process. Further, we provide examples that show it's benefits on a few uses cases. Finally we explore variations of exact risk budgeting that relax the risk budgeting equality constraints to provide long-term economic value to the portfolio.

\subsection{Generalized Risk Budgeting}
As discussed earlier, the risk budgeting problem that can be solved by using the KKT conditions for the modified unconstrained problem is a starting point for the risk budgeting approach. However, two major disadvantages of the formulation are - (a) Asset managers are known to have return forecasts, even though the estimates may be error-prone. (b) Risk budgets are not exactly known, however, the minimum or maximum risk allocation for each asset class usually are known to a sophisticated asset manager.

In particular, we focus on a specific version of this problem where asset managers have strong opinions on the maximum risk budgets they would like to allocate for each asset (or strategy). Indeed, an effective asset allocation strategy could begin with a clear picture of a `max-risk-budget' that a manager would like to have in order to meet return expectations. These could be driven out of sector specific maximum risk allocations or other requirements from an allocation. For example, one can have a desirable maximum allocation to Environmental and Socially Responsible companies (ESG). In this context, we now consider the following {\em mean variance with max-risk budgeting} problem $(\mathbf{\bar{P}})$
\begin{optprog*}\label{eq:rba-prog}
minimize & \objective{- r^\top x + \lambda \,\sqrt{x^\top Cx}}\\ 
$(\mathbf{\bar{P}})$ \qquad subject to & x_i(Cx)_i & \leq & b_i\,x^\top Cx, & i\in[n] \\ 
& l \leq & x & \leq u\\ 
& 1^\top  x & = & 1
\end{optprog*}
where $r_i$ is the return of asset $i$, $\lambda > 0$ controls the mean-variance trade-off, the scalar $b_i$ is the minimum risk budget for asset $i$ and the vectors $l,u\in\R^n_+$ are bounds on portfolio weights. Arguably, this problem does away with all of the limitations of the previous formulation. Return forecasts can be incorporated, upper limits on risk budgets can be set, and position limits are also incorporated. Transaction costs for a multi-period setting can be added with an extra term in the objective function, which we show later.

One observation about this formulation is the following: Suppose there is a feasible solution to $(\mathbf{\bar{P}})$ and the risk budgeting constraints \eqref{eq:par-def} do not hold with equality for all $i$, then, by definition $\exists\, m \in [n]$ such that
\begin{align}\label{eq:rba-sum_contra}
 x_m(Cx)_m < b_m\,x^\top Cx 
\end{align}
where the inequality is strict. Summing across all $i$,
\begin{align}
 \sum_i x_i(Cx)_i < \sum_i b_i\,x^\top Cx
\end{align}
However this suggests that the sum of all marginal risk budgets, which must sum to the total risk, is strictly less than the total risk establishing a contradiction. A similar argument can be made if the strict inequality in \eqref{eq:rba-sum_contra} is facing the other direction.
This implies that risk-budget constraints in $(\mathbf{\bar{P}})$ must hold with equality if the solution exists. We can thus reformulate $(\mathbf{\bar{P}})$ equivalently as a { \em min-risk} constraint risk budgeting problem:
\begin{optprog*}\label{eq:rba-prog-flipped}
minimize & \objective{- r^\top x + \lambda \,\sqrt{x^\top Cx}}\\ 
$(\mathbf{\bar{P}})$ \qquad subject to & x_i(Cx)_i & \geq & b_i\,x^\top Cx, & i\in [n] \\ 
& l \leq & x & \leq u\\ 
& 1^\top  x & = & 1
\end{optprog*}
\section{A Second Order Cone Programming Formulation}
Assume that $x_i$ is non-zero, and satisfies
\begin{align}\label{eq:rba-prog2}
 x_i(Cx)_i \geq  b_i\,x^\top Cx, \,\, i=1,\ldots,n 
\end{align}
Let $Cx = y$ and $C = R^\top R$ be the Cholesky decomposition of $C$. Letting $s_i = \sqrt{b_i} Rx$ we have,
\begin{align}\label{eq:rba-prog3}
 x_i y_i \geq s_i^\top s_i, \,\, i\in [n]. 
\end{align}
\eqref{eq:rba-prog3} represents a rotated second order cone constraint. This can further be reformulated as a second order cone constraint, i.e., for all $i\in [n]$ 
\begin{align}\label{eq:rba-prog5}
    \begin{Vmatrix} x_i - y_i \\ 2s_i \\ \end{Vmatrix} \ \le x_i + y_i 
\end{align}
Alternatively, substituting for $y$ and $s$ we have:
\begin{align}\label{eq:rba-prog6}
    \begin{Vmatrix} x_i - (C x)_i \\ 2 \sqrt{b_i} R x \\ \end{Vmatrix} \ \le x_i + (C x)_i.
\end{align}
Consequently, we can re-write the mean variance optimization problem under risk budgeting constraints in Eq.~(\ref{eq:rba-prog-flipped}) as a convex second order cone program:

\begin{optprog*}\label{eq:rba-socp}
minimize & \objective{-r^\top x + \lambda \, \|Rx\|+ \mu \, \|x-x_0\|_1  }\\ 
subject to & \begin{Vmatrix}
                    x_i - (C x)_i \\ 
                    2 \sqrt{b_i} R x 
            \end{Vmatrix} \  & \leq &  x_i + (C x)_i \\ 
& l \leq & x & \leq u\\ 
& 1^\top  x & = & 1
\end{optprog*}


The above formulation has obvious advantages when it comes to solving this problem at scale \cite{boyd2004convex}. Further more, in the limiting case when $b_i$ are set to $1/m$ where m is the number of assets, it yields the risk parity portfolio. We have included a term for transaction costs that scales as per the L1 norm of the difference between the target position and the previous position in the portfolio. This formulation compared with the previous one as in $(\mathbf{P^*})$ can be tweaked to suit individual portfolio managers requirements.

\section{EMPIRICAL ANALYSIS}
We evaluate the performance of our solution approach on four use cases in this section. Each use case has three flavors of risk budgeting style allocation. The first is a vanilla scenario where the outcome of the min-risk budgeting problem is used as is (MRB portfolio), the second where the output of the optimization problem is used to estimate portfolio expected returns and we liquidate the portfolio (abstain option) if the expected return is less than zero (MRBA portfolio), and the third is a combination of the abstain option with leverage (MRBAL portfolio). The leverage was calculated as the ratio between the ex-ante estimated standard deviation of the CRB portfolio and the MRB portfolio, and the MRBA portfolio was scaled with this ratio to obtain the weights for the MRBAL portfolio. For practical considerations, the permissible leverage for MRBAL was capped at 1.5x.

\subsection{Two Assets}
\label{ex1}
In this example, we show how min-risk budgeting can be combined with a simple momentum forecast to create portfolio value over the long term. We used a 40-10 allocation between the S\&P500 (Ticker: SPY) and iShares Core U.S. Aggregate Bond (Ticker: AGG) ETFs, which involves setting a min-risk budget of 40\% for equity and 10\% for bond markets. Note that these don't have to add up to 100\%; the only condition is that they have to sum to a number less than equal to 100\%. No transaction costs were incorporated. The benchmark used is a equal weighted constant re-balance portfolio (CRB portfolio) that allocates half of the portfolio to the equity and the remaining half to the bond ETF. Both are re-balanced weekly and the weekly median price was used for entry and exit. The mean-variance trade-off parameter $\lambda$ was set at 1. Our estimates of the return in the upcoming week were always set to the returns for the previous week.

We retrospectively let the prices evolve over the 720 weeks (starting on and ending on July 24, 2022) and test the performance of the allocation algorithms over this period (backtest). 

Results are shown in Figure 1. Return forecasts can be unreliable in general, but the results show that having clear ideas on the risk diversification ex-ante is useful, as the risk adjusted returns are higher for the MRB portfolio. The MRBA (min-risk budgeting with abstain option) shows a better return profile with smaller draw-downs. The MRBAL portfolio that stays out of markets when expected return is negative and uses leverage (upto 1.5X) out-performs the constant re-balance portfolio both in terms of return and draw-down profiles.

\begin{figure}[ht]
\includegraphics[scale=0.35]{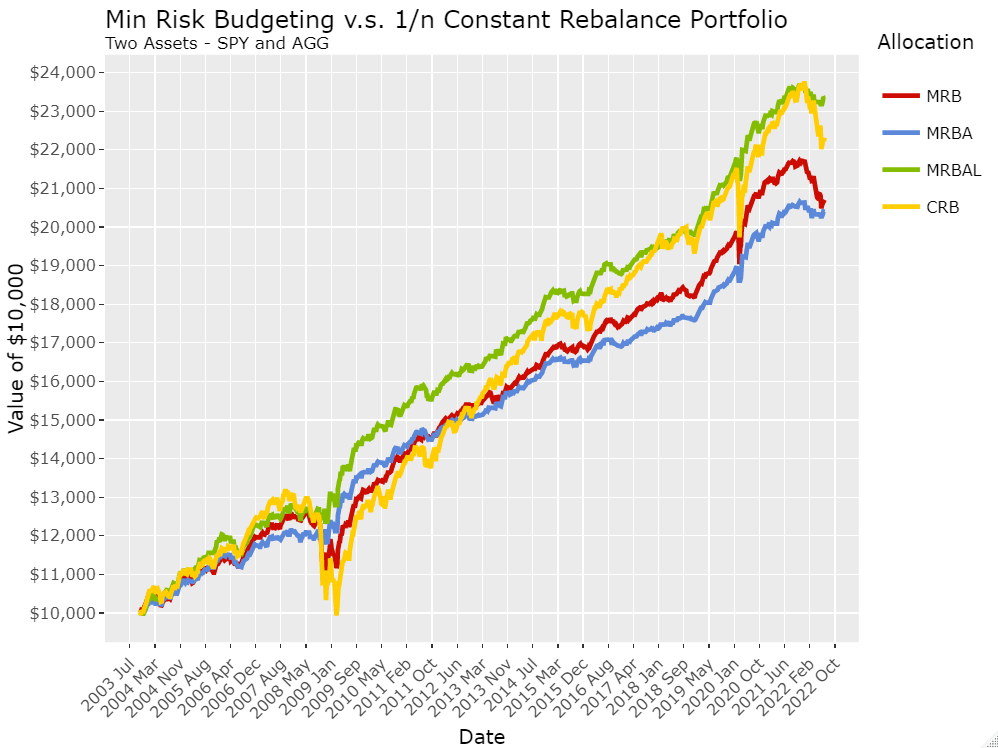}
\caption{Backtest Results: Min Risk Allocation for two assets, SPY 40\%, AGG 10\% }
\end{figure}

Top five drawdowns (absolute value) with dates in focus for the constant re-balance portfolio are shown in Table 1. For all four strategies drawdowns are shown below in Table 2, where the top five drawdowns for each allocation strategy are displayed.

\begin{table}[!htbp] 
\small
\centering 
  \caption{Periods of Drawdowns for CRB portfolio} 
  \label{} 
\begin{tabular}{@{\extracolsep{5pt}} cccc} 
\\[-1.8ex]\hline 
\hline \\[-1.8ex] 
 & From & To & Drawdown.CRB \\ 
\hline \\[-1.8ex] 
1 & 2007-10-19 & 2010-10-08 & $ 0.287$ \\ 
2 & 2020-02-28 & 2020-07-17 & $0.169$ \\ 
3 & 2022-01-07 & 2022-07-24  & $0.164$ \\ 
4 & 2018-09-07 & 2019-03-15 & $0.068$ \\ 
5 & 2011-07-15 & 2012-01-06 & $0.057$ \\ 
\hline \\[-1.8ex]
\end{tabular} 
\label{dd1table}
\end{table} 

\begin{table}[!htbp] 
\small
\centering 
  \caption{Top five drawdowns for each portfolio} 
\begin{tabular}{@{\extracolsep{5pt}} ccccc} 
\\[-1.8ex]\hline 
\hline \\[-1.8ex] 
 & DD.CRB & DD.MRB & DD.MRBA & DD.MRBAL \\ 
\hline \\[-1.8ex] 
1 & 0.29 & 0.15 & 0.04 & 0.05 \\ 
2 & 0.17 & 0.12 & 0.04 & 0.05 \\ 
3 & 0.16 & 0.08 & 0.04 & 0.05 \\ 
4 & 0.07 & 0.05 & 0.04 & 0.05 \\ 
5 & 0.06 & 0.03 & 0.03 & 0.04 \\ 
\hline \\[-1.8ex]
\label{dd2table}
\end{tabular}
  \label{ex1drawdowns} 

\end{table}

The CRB portfolio has a staggering max drawdown of 29 \% compared to the MRBA portfolio at 4 \%.

\subsection{Four Assets, including Gold and SPY Levered ETF}
In this example, we repeat the experiment in \ref{ex1} with an expanded set of assets and we change the benchmark to a risk parity allocation. In particular, we include a commodity and a leveraged ETF to enhance leverage and diversity in the portfolio. The two additional ETFs chosen were SPXL (Direxion Daily S\&P500 Bull 3X) and GLD (SPDR Gold Shares). As a result of limited history for SPXL, we have 3,451 days of data which translate to 720 weeks of data as of July 24 2022. The chosen risk budget allocations across SPY, AGG, SPXL, GLD were 35\%, 5\%, 35\%, 5\%. The risk parity portfolio has these set to 25\% risk for each asset. Other parameters were kept the same as \ref{ex1}.

\begin{figure}[h]
\includegraphics[scale=0.35]{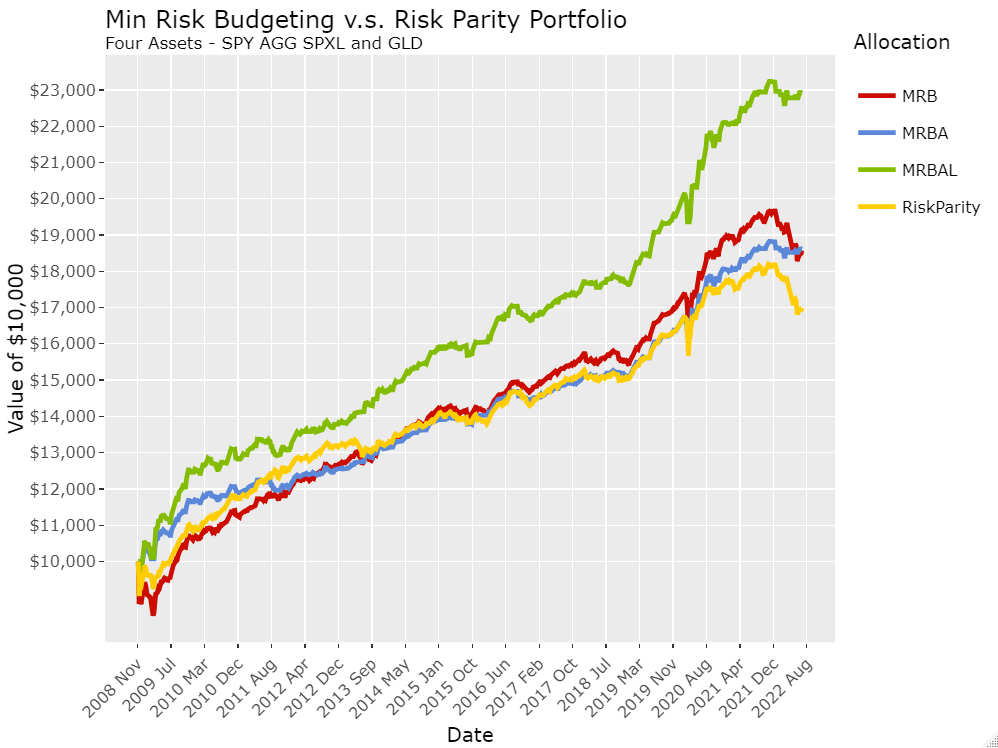}
\caption{Backtest Results: Min Risk Allocation at SPY 35\%, AGG 5\%, SPXL 35\%, GLD 5\%}
\end{figure}

The results show that the Risk Parity strategy can be adjusted with a flexible parameter of a min-risk setting, and the allocation that comes from the framework enhances the return profile in case of both the leveraged (MRBAL) and unleveraged (MRB, MRBA) version of the backtest. 


\subsection{NASDAQ 100 Constituents}
\label{ex3}
In this example we used a selection of 64 assets (listed cash equity) from the US stock market that are a subset of the NASDAQ 100 index constituents. Tickers that were part of the index as of July 24, 2022 and had availability of daily price time-series data for at least 5,423 (a sufficiently large number) trading days were chosen as the asset universe. The tickers were retrieved from the NASDAQ web page \cite{nasdaqWebsite} 
on the same date, 102 tickers were filtered using the above criteria to yield 64 tickers with at least the minimum threshold of price history. Daily price data was downloaded from EODHistoricalData's (a commercial service) end of day API \cite{eodWebsite}
. We down-sampled the daily price data to a weekly time-series and used the weekly median price as the entry and exit price (or vice versa) for the trading simulation. We used momentum return forecasts by setting forecast return to be the same as the previous week's return. The minimum risk budget for each asset is set as half of the cross-sectionally normalized cumulative return thus far for that asset. The ratio of half can be chosen arbitrarily, the key concept here is that we are choosing less than the desired allocation as a lower limit on the risk budget to allow trade offs between the return forecast and the risk budget. The covariance matrix is measured point in time including all of the returns thus far and if not already positive-definite, it was adjusted to obtain the nearest positive-definite matrix following the result in \cite{higham1988computing} that the nearest symmetric positive semi-definite matrix in the Frobenius norm to an arbitrary real matrix $C$ is $(B + H)/2$, where H is the symmetric polar factor of $B=(C + C')/2.$

\begin{figure}[h]
\includegraphics[scale=0.35]{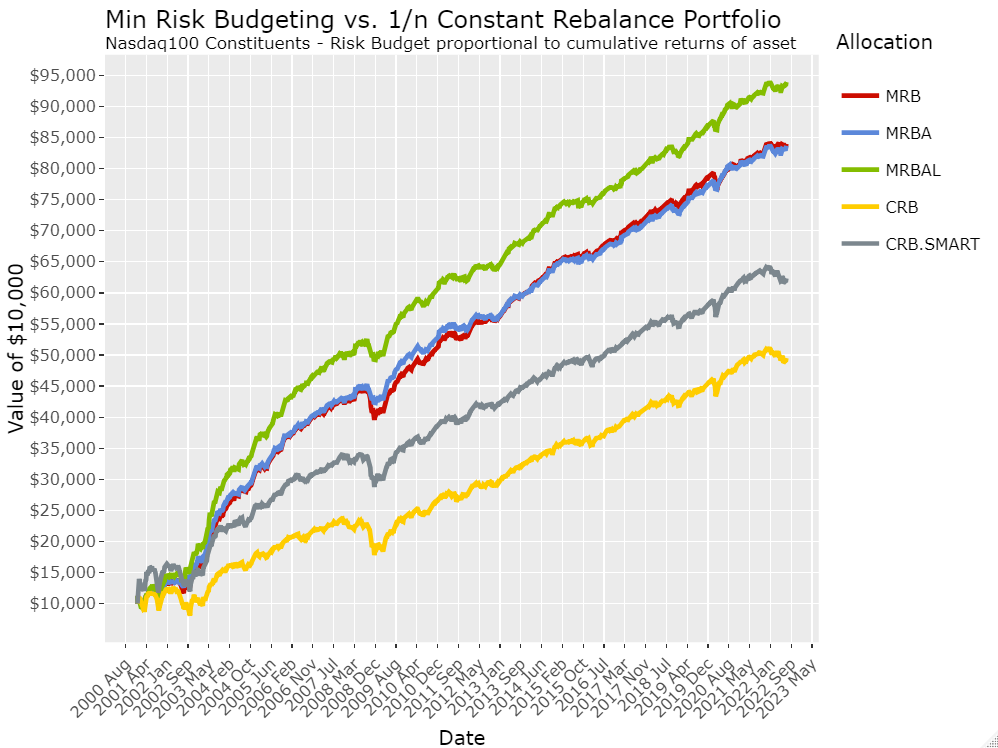}
\caption{Backtest Results: Allocation into NASDAQ 100 constituents}
\label{nasdaq}
\end{figure}

The results are shown in Figure \ref{nasdaq}. We also introduce a new benchmark that sets dollar weighting of the asset to be the cross-sectionally normalized cumulative return thus far for that asset, in order to produce a benchmark that is closer to the MRB back tests. We denote this second benchmark as CRB.SMART and it is shown in comparison to the first benchmark CRB and the min risk budget backtests (MRB, MRBA and MRBL). Note that this backtest has an inherent look head bias in that it uses assets that survive the period between September 2000 and July 2022. Hence the asset universe comprises of higher quality assets than the NASDAQ at that point in time. Since we are comparing only the relative performance of the class of MRB* portfolios to the risk parity portfolio, this chart serves its purpose as a relative comparison and is not indicative of absolute performance.

Drawdown tables are shown in Table \ref{nasdaqtable}.

\begin{table}[!htbp] 
\tiny
\centering 
  \caption{Top five drawdowns for each portfolio - NASDAQ Constituents} 
   
\begin{tabular}{@{\extracolsep{5pt}} cccccc} 
\\[-1.8ex]\hline 
\hline \\[-1.8ex] 
 & CRB & CRB.SMART & MRB & MRBA & MRBAL \\ 
\hline \\[-1.8ex] 
1 & 0.46 & 0.44 & 0.42 & 0.26 & 0.28 \\ 
2 & 0.38 & 0.38 & 0.26 & 0.18 & 0.26 \\ 
3 & 0.27 & 0.26 & 0.21 & 0.14 & 0.18 \\ 
4 & 0.26 & 0.25 & 0.19 & 0.14 & 0.17 \\ 
5 & 0.23 & 0.18 & 0.13 & 0.14 & 0.17 \\ 
\hline \\[-1.8ex] 
\end{tabular} 
\label{nasdaqtable}

\end{table}

\subsection{FTSE 100 Constituents}
\begin{figure}[h]
\includegraphics[scale=0.35]{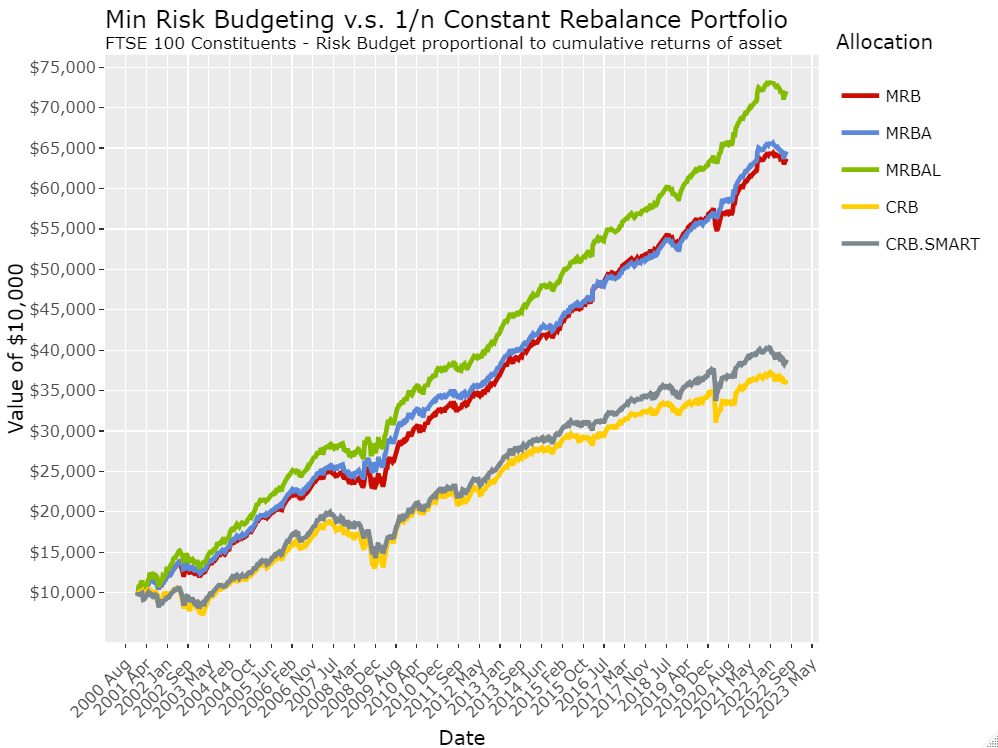}
\caption{Backtest Results: Allocation into FTSE 100 constituents}
\label{ftse}
\end{figure}

Here we use a European equity market selection of 72 assets (listed cash equity) from the stocks that are listed on the London Stock Exchange(LSE) and are part of the FTSE 100 as of July 24 2022. The criteria for selection was the same as in the previous example in \ref{ex3}. Tickers that were part of the FTSE index as of July 24, 2022 and had availability of daily price time-series data for 5,448 trading days were chosen as the asset universe.
The tickers were retrieved from the LSE web page \cite{ftseWebsite} 
on the same date, 102 tickers were filtered using the criteria mentioned above to yield 72 tickers. Trading simulation parameters are same in \ref{ex3}, in addition a 10 basis point transaction cost parameter was incorporated. The results are shown in Figure \ref{ftse}. The results confirm that setting minimum values of risk budget can be an important tool in an investment professional's arsenal.
Drawdown tables are shown in Table \ref{ftsetable}.

\begin{table}[!htbp] 
\tiny
\centering 
\caption{Top five drawdowns for each portfolio - FTSE Constituents}
\begin{tabular}{@{\extracolsep{5pt}} cccccc} 
\\[-1.8ex]\hline 
\hline \\[-1.8ex] 
 & CRB & SMART & MRB & MRBA & MRBAL \\ 
\hline \\[-1.8ex] 
1 & 0.47 & 0.45 & 0.26 & 0.17 & 0.19 \\ 
2 & 0.33 & 0.35 & 0.25 & 0.17 & 0.19 \\ 
3 & 0.30 & 0.23 & 0.20 & 0.16 & 0.18 \\ 
4 & 0.20 & 0.19 & 0.18 & 0.15 & 0.17 \\ 
5 & 0.17 & 0.18 & 0.14 & 0.14 & 0.16 \\ 
\hline \\[-1.8ex] 
\end{tabular} 
\label{ftsetable}
\end{table} 

\section{CONCLUSION}

Portfolio optimization is an area where techniques from operations research can be used to create more stable portfolios, keeping other factors the same. We have shown a powerful technique for portfolio optimization where risk and return can be traded off systematically to create resilient portfolios. Due to the reliance of the allocation strategy on variance information only, these methods are applicable whenever price data is available, making these techniques very flexible. The technique described in this work can be useful for sophistical retail investors, traditional investment management companies and also to the more new-age robo-advisors. 

\bibliography{ERB_Bibliography}

\end{document}